\documentclass[runningheads]{llncs}
\usepackage[T1]{fontenc}
\usepackage{graphicx}
\usepackage{subfig}
\usepackage{cite}
\usepackage{hyperref}
\usepackage{booktabs}
\usepackage{amsmath}
\begin{document}
\title{Enhancing Quantum Diffusion Models with Pairwise Bell State Entanglement}
\titlerunning{EEQDM}
%
\author{Shivalee RK Shah \and Mayank Vatsa}
\authorrunning{S. Shah and M. Vatsa}
%
\institute{Indian Institute of Technology Jodhpur, Rajasthan, India\\
\email{shah.14@iitj.ac.in, mvatsa@iitj.ac.in}
}
\maketitle              
\begin{abstract}
This paper introduces a novel quantum diffusion model designed for Noisy Intermediate-Scale Quantum (NISQ) devices. Unlike previous methods, this model efficiently processes higher-dimensional images with complex pixel structures, even on qubit-limited platforms. This is accomplished through a pairwise Bell-state entangling technique, which reduces space complexity. Additionally, parameterized quantum circuits enable the generation of quantum states with minimal parameters, while still delivering high performance. We conduct comprehensive experiments, comparing the proposed model with both classical and quantum techniques using datasets such as MNIST and CIFAR-10. The results show significant improvements in computational efficiency and performance metrics such as FID, SSIM and PSNR. By leveraging quantum entanglement and superposition, this approach advances quantum generative learning. This advancement paves the way for more sophisticated and resource-efficient quantum diffusion algorithms capable of handling complex data on the NISQ devices.

\end{abstract}
\section{Introduction}

Quantum computing has seen remarkable progress in recent years, opening up new possibilities for solving intricate computational challenges. Specifically, in the field of image generation and machine learning, Quantum Denoising Diffusion Models (QDDMs) are emerging as a promising technology to enhance both the efficiency and effectiveness of these applications. While traditional (non-quantum) diffusion models are quite capable, they often require extensive parameter tuning and can be computationally demanding \cite{dhariwal2021diffusion, ho2020denoising, rombach2021high}.

Diffusion models gradually transform a simple noise distribution into a complex data distribution through a series of iterative steps. This procedure is inherently computationally intensive, especially as the size and and complexity of training dataset grow. Quantum diffusion models, however, capitalize on the unique properties of quantum mechanics—namely superposition and entanglement—to circumvent these challenges\cite{biamonte2017quantum, gabor2020holy}. Quantum entanglement facilitates the creation of highly correlated states, which can be efficiently manipulated to perform complex transformations, while superposition permits quantum bits (qubits) to occupy multiple states at once, dramatically enlarging the computational space. These quantum characteristics make diffusion models especially powerful for generative tasks involving large datasets and complex, high-dimensional data.

\begin{figure*}[t]

    \centering
    \includegraphics[width=\textwidth]{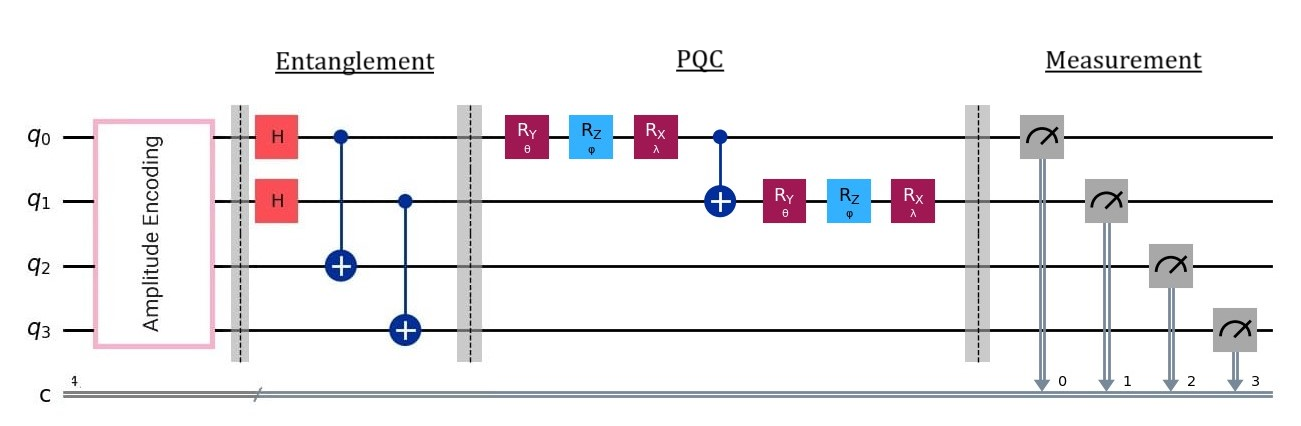}
    \caption{Schematic overview of the proposed Entanglement Enhanced Quantum Diffusion Model (EEQDM). The circuit illustrates the key components of EEQDM: (1) Amplitude encoding for initial state preparation, (2) Entanglement generation using Hadamard ($H$) and CNOT gates, enhancing the model's capability, (3) Parameterized Quantum Circuit (PQC) layer with rotation gates $R_y(\theta)$, $R_z(\phi)$, and $R_x(\lambda)$ and CNOT gate, implementing the diffusion process, and (4) Measurement stage. This architecture leverages pairwise entanglement to enhance the quantum diffusion process, potentially improving performance in tasks such as generative modeling or optimization. Qubits $q_0$-$q_3$ represent the quantum register, while $c$ denotes the classical measurement outcomes.}
    \label{fig:comparison}
\end{figure*}


As shown in Fig. \ref{fig:comparison}, this paper proposes Entanglement Enhanced Quantum Diffusion Model (EEQDM) architecture that leverages the quantum properties through a carefully designed circuit. The circuit begins with amplitude encoding, followed by an entanglement stage using Hadamard (H) gates and CNOT operations. This entanglement facilitates the creation of highly correlated \textbf{Bell state pairs}, which can be efficiently manipulated to perform complex transformations. The key feature of the approach is the application of a Parameterized Quantum Circuit (PQC) to only a subset of qubits (q0 and q1 in the image). This selective application of the PQC significantly reduces the parameter count while maintaining the power of quantum processing. The circuit concludes with a measurement stage, allowing to extract the processed information. The proposed method is evaluated against traditional classical models and current quantum models using widely recognized datasets like MNIST digits and CIFAR-10, which are frequently utilized in quantum machine learning research. The findings show an increase in computational efficiency and performance metrics, highlighting the effectiveness of the entanglement strategy employed in the proposed model to enhance QDDMs.

\section{Literature Review and Related Work}

Recent advancements in quantum machine learning have highlighted the potential of Quantum Diffusion Models (QDMs) in improving image generation tasks. Classical diffusion models, such as Denoising Diffusion Probabilistic Models (DDPMs), have been instrumental in advancing image synthesis but are often hampered by high computational demands and the need for extensive parameter tuning \cite{dhariwal2021diffusion}. The transition to quantum-based models offers a promising solution to these challenges.

\subsection{Classical Diffusion Models}

Classical diffusion models are generative models designed to learn the probability distribution \(p(x)\) of a dataset, enabling the generation of new samples from this distribution. The diffusion process involves a Markov chain that gradually maps an arbitrary distribution \(q(x_0)\) to a simpler, treatable distribution \(\pi(x_T)\), often a Gaussian distribution. This is achieved through a forward process using a Markov kernel \(q(x_t|x_{t-1})\), and a parametric model is trained to estimate the inverse Markov chain, \(p_\theta(x_{t-1}|x_t)\) \cite{dhariwal2021diffusion}.

\subsection{Parameterized Quantum Circuits}

Parameterized Quantum Circuits (PQCs) are essential components in the domain of quantum machine learning\cite{schuld2015introduction}, serving as the quantum analog of classical artificial neural networks. PQCs are composed of quantum gates that perform parametric transformations on quantum states, organized in layers to facilitate complex data processing. Each quantum gate within these circuits is defined by rotation angles, which are trainable parameters optimized using techniques such as gradient descent \cite{koelle2024quantum}. The strong entangling ansatz, which combines trainable rotation gates (\(R_x\), \(R_y\), \(R_z\)) with C-NOT gates to create entanglement between qubits, has been particularly effective \cite{kim2023quantum}.

Training PQCs on current noisy intermediate-scale quantum (NISQ) \cite{preskill2018quantum} hardware poses challenges due to high noise levels. Consequently, simulations using software libraries like PennyLane are often employed for training, where both forward computations and optimizations are performed on classical computers. These simulations encode classical data into quantum states using amplitude encoding, which maps a classical vector's components onto the coefficients of a quantum state, allowing the representation of \(2^N\) classical features with \(N\) qubits \cite{koelle2023enhancing}. However, this encoding requires a number of C-NOT gates that grows exponentially with the number of qubits, presenting a scalability challenge.

\subsection{Quantum Diffusion Models}

Quantum Diffusion Models (QDMs) combine the principles of quantum computing with the methodology of diffusion models to enhance generative capabilities. These models leverage quantum mechanics properties such as superposition and entanglement to process information more efficiently than classical models. Quantum Denoising Diffusion Models (QDDMs) utilize parameterized quantum circuits to model the data distribution, providing a compact representation that reduces computational load compared to classical counterparts \cite{koelle2024quantum}.

A significant development in QDMs is the use of intermediate measurements and ancillary qubits, which have been shown to improve the quality of generated samples by introducing non-linear mappings over state amplitudes \cite{koelle2024quantum}. However, excessive measurements can lead to model collapse due to the loss of initial noise information. Hybrid models, combining classical autoencoders with QDMs, have also demonstrated enhanced performance by simplifying PQCs and enabling implementation on real quantum hardware \cite{koelle2023enhancing}. These latent models reduce the dimensions of the input data before processing with quantum circuits, improving efficiency but adding complexity by requiring classical pre-processing. 

The proposed research further enhances this approach by reducing the parameter count directly within the quantum circuit itself, eliminating the need for classical autoencoders. The Bell state entanglement strategy helps in maintaining the fidelity of quantum states across iterations, reducing errors, and enabling the model to handle datasets with fewer parameters. By benchmarking EEQDM against classical and existing quantum models using datasets like MNIST digits and CIFAR-10, we demonstrate significant improvements in computational efficiency and performance metrics, showcasing the potential of our approach in enhancing QDMs.

\section{Methodology}

In this section, we provide the details of the methodology employed to construct EEQDM. The proposed approach integrates the design of a quantum variational circuit with the implementation of a diffusion process enhanced by an entanglement-based technique.

\subsection{Construction of Quantum Diffusion Models}

The Entanglement-Enhanced Quantum Diffusion Model (EEQDM) introduces a novel quantum circuit design that harnesses the power of Bell-state entanglement to enhance performance while simultaneously reducing the parameter count in quantum diffusion models. The architecture of EEQDM consists of three primary stages: Amplitude Encoding, Pair-wise Bell-state preparation, and Parameterized Quantum Circuit (PQC), followed by a measurement stage.


\subsubsection{Input: Amplitude Encoding}
We begin by performing amplitude encoding to embed the input data into the quantum circuit. This method uses \(\log(n)\) qubits, where, \(n\) is the number of features in the dataset. Amplitude encoding is efficient for handling high-dimensional data as it maps the classical data vector components onto the amplitude of quantum states. let \(|x\rangle\) be the quantum state representing the input data vector \(x\). Mathematically, amplitude encoding can be expressed as:
\begin{equation}
    |x\rangle = \sum_{i=0}^{n-1} x_i |i\rangle
\end{equation}

We first flatten the 2D image data into a 1D vector, normalize it, and then apply amplitude encoding. This process allows us to represent an image of N pixels using only $(\lceil\log_2(N)\rceil)$ qubits \cite{mottonen2004transformation, steane1998quantum}, significantly reducing the quantum resources required compared to direct qubit encoding methods. For instance, a $16\times16$ image (256 pixels) would require only 8 qubits, while a $64\times64$ image (4096 pixels) would need 12 qubits. This logarithmic scaling in qubit requirements demonstrates the efficiency of amplitude encoding for handling various image sizes in EEQDM.

\subsubsection{Entanglement}
Following the amplitude encoding, the Pair-wise Bell-state preparation stage implements a specific entanglement strategy that creates a unique quantum state, efficiently distributing information across the qubits. This process can be described as "pairwise qubit entanglement" and unfolds as follows:
\begin{enumerate}
    \item Hadamard gates (H) are applied to the first half of the qubits (excluding the ancilla qubit). This creates an equal superposition state for these qubits, preparing them for entanglement.
    \item CNOT gates are then used to entangle each qubit from the first half with a corresponding qubit in the second half. Specifically, for n qubits (excluding the ancilla), we apply CNOT gates where, the control qubits are from the first half (indices 0 to n/2-1) and the target qubits are their corresponding partners in the second half (indices n/2 to n-1).
\end{enumerate}
\begin{figure*}[t]

    \centering
    \includegraphics[width=\textwidth]{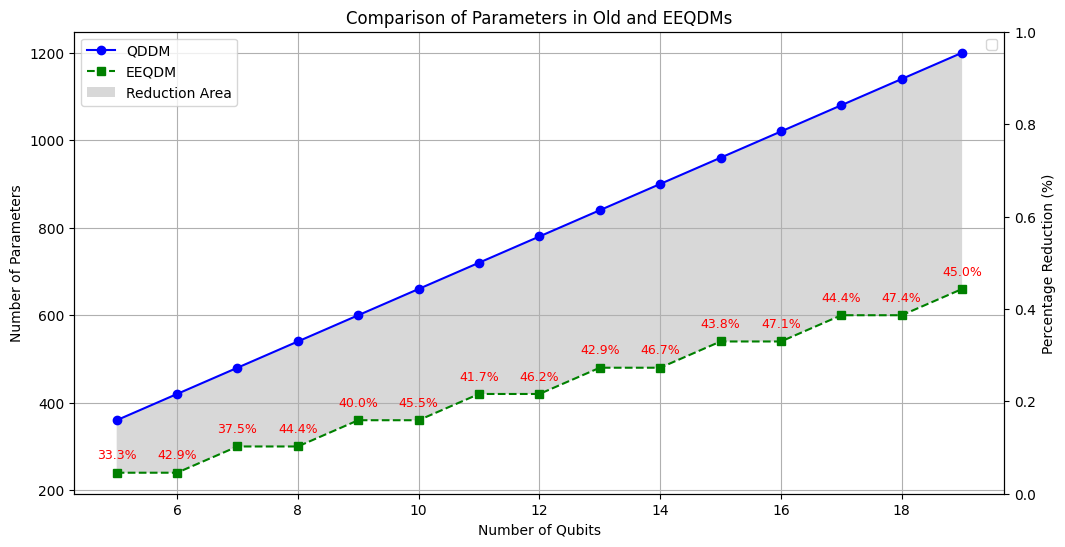}
    \caption{Comparison of parameters in QDDM\cite{koelle2024quantum} and the proposed EEQDM as the number of qubits increases. The shaded area highlights the parameter reduction achieved by the new model.}
    \label{fig:comparison_params}
\end{figure*}



The Pair-wise Bell-state preparation is crucial to EEQDM's parameter reduction mechanism. By creating specific entanglement pairs between the first and second half of the qubit register, we establish information pathways that allow the subsequent Parameterized Quantum Circuit (PQC) to operate on a reduced set of qubits while still accessing information from the entire input state. Thus, significantly reducing the number of parameters required while maintaining the model's expressive power.

The effectiveness of this approach is evident in the parameter reduction graph Fig. \ref{fig:comparison_params}, which shows a consistent 40-47\% reduction in parameters for qubit counts between 8 and 18, compared to existing quantum diffusion models. This substantial reduction in parameters, enabled by the unique entanglement strategy, is a key factor in EEQDM's improved efficiency and scalability.
\subsubsection{Ansatz: Parameterized Quantum Circuit}
The foundational element of EE-QDM architecture is a Parameterized Quantum Circuit (PQC) which serves as the ansatz. We conduct an extensive exploration of various circuit depths and entanglement architectures within these PQCs to ascertain the optimal configuration for the proposed quantum diffusion model. Each layer of the ansatz is composed of trainable parameters, facilitated through rotation and C-NOT gates, enabling the circuit to dynamically adapt and learn from the data throughout the training process \cite{Benedetti2019}. In our experiments, we vary the depth (L) and entanglement patterns within the PQCs to determine the configuration that best aligns with the requirements of the diffusion model \cite{Cerezo2021}. This framework permits the circuit to effectively adapt to the data distribution across different noise levels encountered during the diffusion process. The selection of this ansatz is strategically motivated by the need to balance expressivity with parameter efficiency, which is essential for streamlined training and to prevent overfitting \cite{Sim2019}. This balance is crucial for optimizing the learning capabilities of our model while ensuring generalization across diverse datasets \cite{Kandala2017}.

\begin{figure}[h]
    \centering
    \includegraphics[width=0.5\textwidth]{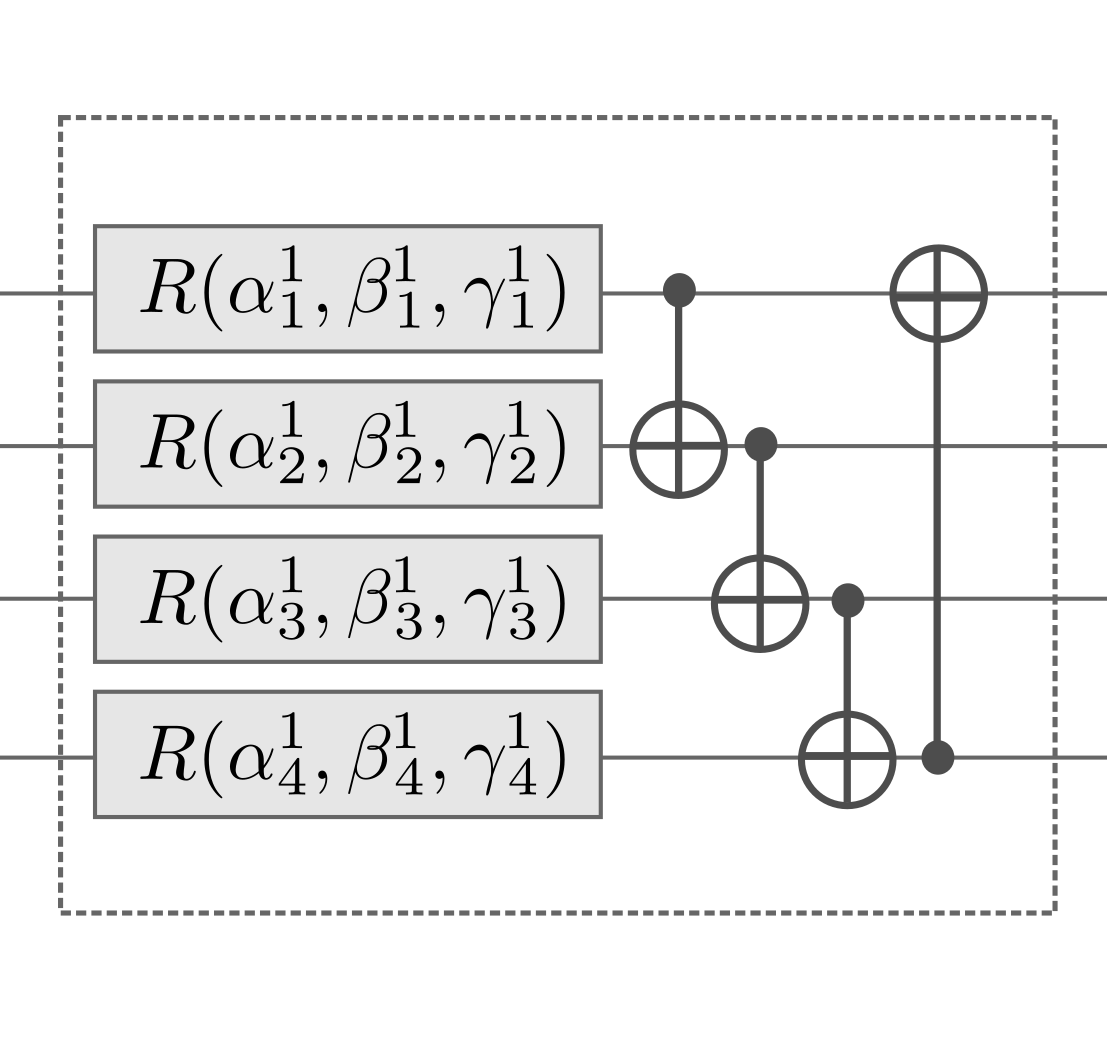}
    \caption{Example of ansatz with 4-qubit strongly entangling layers ($L = 1$) showing rotations $R$ and CNOTs. In practice, we use multiple layers and optimize the depth for our specific task. Image from PennyLane documentation, available at: \href{https://docs.pennylane.ai/en/stable/code/api/pennylane.StronglyEntanglingLayers.html}{PennyLane Documentation}.}
    \label{fig:entangling_layers}
\end{figure}

\subsection{Diffusion Process}

The diffusion process follows a Markov chain framework, where, data undergoes forward diffusion to introduce noise, followed by reverse diffusion to denoise and reconstruct the data. This process is inspired by classical diffusion models but adapted to the quantum domain.
\begin{figure}

    \centering
    \includegraphics[width=\textwidth]{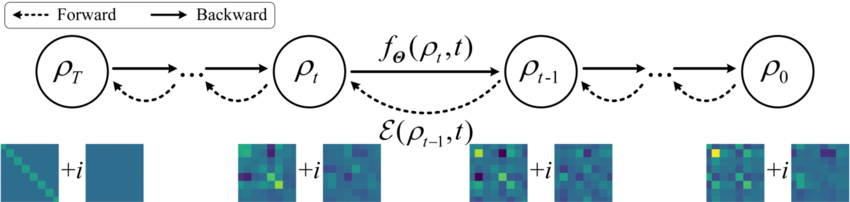}
    \caption{It depicts the state evolution from $\rho_T$ to $\rho_0$ through intermediate states $\rho_t$ and $\rho_{t-1}$. Each step involves applying the function $\mathcal{E}(\rho_{t-1}, t)$and the quantum operation $f_\theta(\rho, t)$. The bottom row showcases visual representations of these states, highlighting the transformation and diffusion process in the quantum generative model.\cite{chen2024quantum}}
    \label{fig:diffusion process}
\end{figure}
\subsubsection{Forward Diffusion}

Noise is incrementally added to the data across multiple timesteps. The forward diffusion process can be mathematically represented as:
\begin{equation}
    q(x_t|x_{t-1}) = \mathcal{N}(x_t; \sqrt{\alpha_t}x_{t-1}, (1-\alpha_t)I) 
\end{equation}
where, \(\alpha_t\) controls the variance schedule.

\subsubsection{Reverse Diffusion}
The reverse diffusion process, also known as the sampling or generation process, is a crucial component of diffusion models. This process involves gradually denoising a random input to produce a high-quality sample. Our implementation of reverse diffusion is inspired by recent advancements in the field \cite{dhariwal2021diffusion, koelle2024quantum, kim2023quantum}.

The process begins with a random noise tensor \( x_0 \), typically sampled from a standard normal distribution. This noise is then iteratively refined through \( T \) steps, where, \( T \) is the number of diffusion steps:

\begin{equation}
    x_0 \sim \mathcal{N}(0, I)
\end{equation}

\begin{equation}
    x_1, x_2, \ldots, x_T = ReverseProcess(x_0)
\end{equation}

At each step \( t \), the proposed model predicts either the denoised data directly or the noise component, depending on the chosen prediction goal. For data prediction, we directly use the network output:

\begin{equation}
  x_t = f_\theta(x_{t-1})  
\end{equation}

where, \( f_\theta \) is our quantum circuit with parameters \( \theta \).
This direct prediction of denoised data, rather than noise, has empirically shown better results in experiments. Our implementation also supports conditional generation, where, additional label information can be provided to guide the reverse diffusion process. This is particularly useful for tasks requiring controlled generation or class-specific sampling.




\subsection{Training and Optimization}

Training involves optimizing the PQC parameters to minimize the reconstruction error. We use the Mean Squared Error (MSE) loss function to quantify the difference between the reconstructed and original data:
\begin{equation}
     \mathcal{L} = \frac{1}{N} \sum_{i=1}^{N} \| x_i - \hat{x}_i \|^2 
\end{equation}
where, \( x_i \) and \( \hat{x}_i \) represent the original and reconstructed data, respectively.


Our hybrid quantum-classical pipeline integrates quantum computing capabilities with classical optimization techniques to solve complex problems efficiently. At its core, a parameterized quantum circuit processes and encodes input data into quantum states. The circuit's output is measured and classically post-processed to compute the objective function.
The circuit parameters are optimized using the classical Adam optimizer. We conducted hyperparameter tuning for the learning rate, testing values of 0.1, 0.01, and 0.001, with 0.1 yielding the best results. This iterative process of quantum computation followed by classical optimization allows us to harness the potential quantum advantage in data processing while leveraging well-established classical algorithms for parameter updates.
This hybrid approach enables us to tackle problems that may be challenging for purely classical or quantum methods, potentially opening new avenues in optimization and machine learning tasks.

Performance metrics such as Structural Similarity Index Measure (SSIM), and Peak Signal-to-Noise Ratio (PSNR) are employed to evaluate model performance. These metrics assess the model's ability to generate high-quality images that closely resemble the original data distribution.

\section{Experimental Setup}
\subsection{Datasets}
In our experiments, we utilized two standard benchmark datasets widely recognized in both classical and quantum machine learning communities: MNIST \cite{lecun1998gradient} and CIFAR-10\cite{krizhevsky2009cifar} . These datasets were chosen for their widespread use in evaluating image processing models, including recent quantum machine learning approaches \cite{adhikary2024supervised, cao2023quantum, feng2023quantum}.
The MNIST dataset consists of 70,000 grayscale images of handwritten digits ($28\times28$ pixels), split into 60,000 training images and 10,000 test images. This dataset has been extensively used in quantum machine learning literature due to its simplicity and the clear benchmarking it provides for digit recognition tasks \cite{adhikary2024supervised, feng2023quantum}. CIFAR-10 includes 60,000 color images ($32\times32$ pixels) across ten different classes, with 50,000 training images and 10,000 test images. Its inclusion allows us to evaluate our model's performance on more complex, real-world images, following recent trends in quantum image processing research \cite{cao2023quantum, feng2023quantum}.

To assess EEQDM's ability to handle varying data dimensionality and to ensure compatibility with different qubit counts in our quantum system, we preprocessed the images to three different resolutions: $8\times8$, $16\times16$, and $32\times32$ pixels. This approach not only aligns with recent quantum image processing studies \cite{adhikary2024supervised, cao2023quantum} but also challenges the model to manage increasingly larger feature sets effectively, providing insights into its scalability and performance across different data complexities.

\subsection{Metrics}

\subsubsection{Loss Curves}
We analyze the training and validation loss curves to evaluate the model's learning progress and generalization ability. The loss function used is Mean Squared Error (MSE), defined as:
\begin{equation}
    {MSE} = \frac{1}{n} \sum_{i=1}^n (y_i - \hat{y}_i)^2
\end{equation}

where, $y_i$ are the true values and $\hat{y}_i$ are the predicted values.
\subsubsection{Structural Similarity Index (SSIM) \cite{wang2004image}} 
SSIM assesses the perceived quality of generated images compared to the originals:
\begin{equation}
    SSIM(x,y) = \frac{(2\mu_x\mu_y + c_1)(2\sigma_{xy} + c_2)}{(\mu_x^2 + \mu_y^2 + c_1)(\sigma_x^2 + \sigma_y^2 + c_2)}
\end{equation}

where, \(\mu_x\) and \(\mu_y\) are the average pixel intensities, \(\sigma_x\) and \(\sigma_y\) are the standard deviations, and \(\sigma_{xy}\) is the covariance of pixels in images \(x\) and \(y\). SSIM ranges from -1 to 1, with 1 indicating perfect structural similarity.

\subsubsection{Peak Signal-to-Noise Ratio (PSNR)} 
PSNR quantifies the ratio between the maximum possible signal power and the power of distorting noise:
\begin{equation}
    PSNR = 20 \cdot \log_{10}\left(\frac{{MAX}_I}{\sqrt{MSE}}\right)
\end{equation}

where, \({MAX}_I\) is the maximum possible pixel value, and MSE is the mean squared error between the generated and original images. Higher PSNR values generally indicate better reconstruction quality. This comparison helps us understand the learning behavior in both quantum and classical domains, as well as the specific benefits and trade-offs of the proposed quantum diffusion approach.

\subsubsection{Fréchet Inception Distance (FID) \cite{heusel2017gans}} 
FID measures the similarity between the distribution of generated images and that of real images. It is computed as:
\[
\text{FID} = ||\mu_r - \mu_g||^2 + \text{Tr}(\Sigma_r + \Sigma_g - 2(\Sigma_r\Sigma_g)^{1/2})
\]
where \(\mu_r\) and \(\mu_g\) are the mean feature representations, and \(\Sigma_r\) and \(\Sigma_g\) are the covariance matrices for real and generated images, respectively. Lower FID scores indicate higher quality and diversity.








\section{Results}
\subsection{Comparison of EEQDM and QDDM}

\subsubsection{Model Specifications}: The proposed Entanglement-Enhanced Quantum Diffusion Model (EEQDM) incorporates a novel pairwise qubit entanglement technique aimed at improving scalability and efficiency. We compared EEQDM with an existing Quantum Denoising Diffusion Model (QDDM) that does not include this feature. Both models were evaluated using 8×8 and 16×16 MNIST images, as well as 16×16 CIFAR-10 images, across depths ranging from 10 to 50.


\subsubsection{Evaluation of Models}
\begin{figure}
    \centering
    \includegraphics[width=\textwidth]{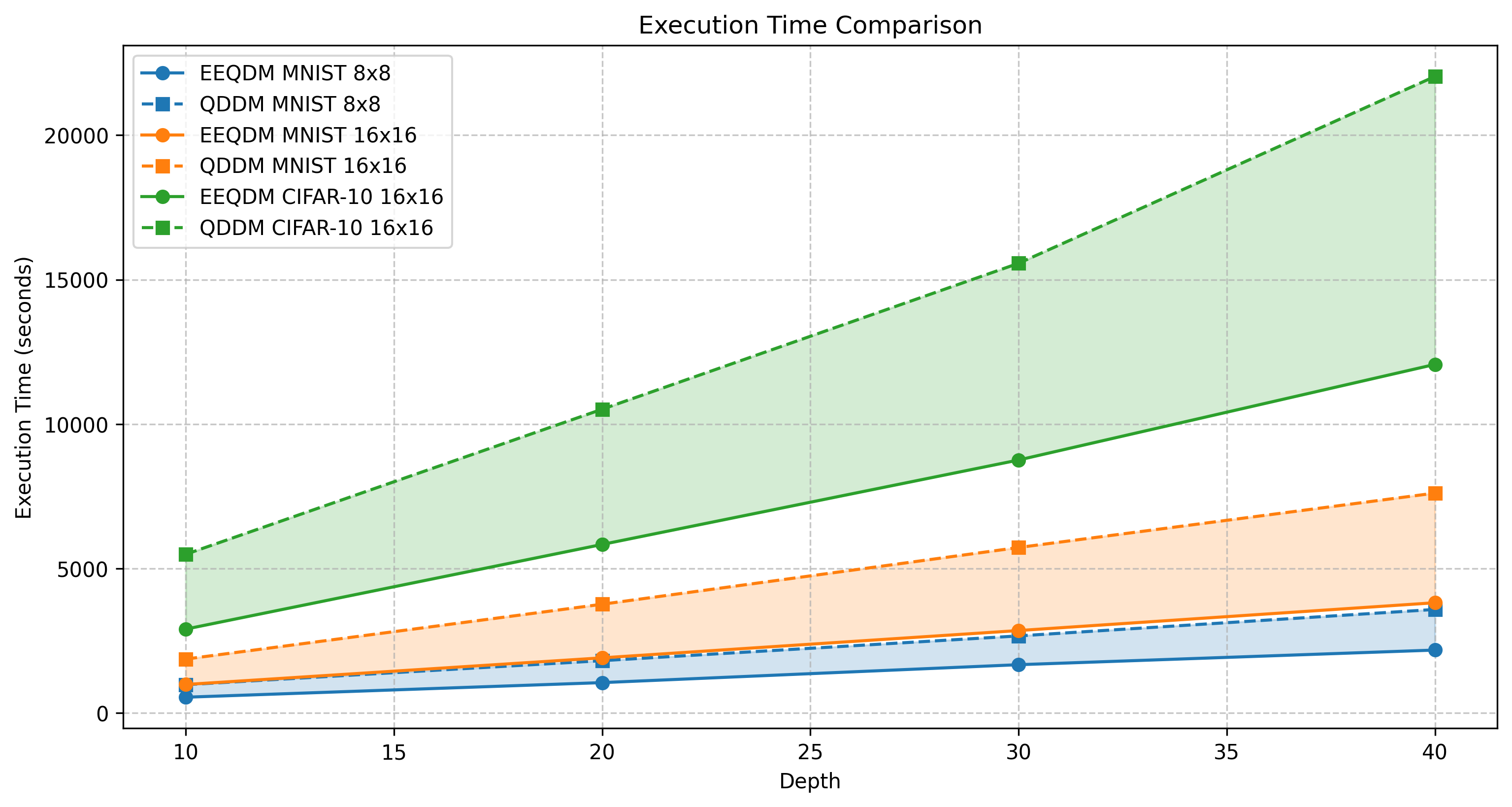}
    \caption{Execution Time Comparison of EEQDM and QDDM for $8\times8$ and $16\times16$ Images Across Various Depths demonstrating that EEQDM consistently outperforms QDDM in terms of execution speed across all depths and configurations, with the performance gap widening as the network depth and data complexity increases.}
    \label{fig:execution_times}
\end{figure}

: As depicted in Fig. \ref{fig:perf_matrix} and Fig. \ref{fig:execution_times}, a detailed comparison of performance metrics and computational efficiency reveals several key trends:





\begin{figure}
\centering

\subfloat[Performance comparison for $8\times8$ MNIST images.]{
	\label{subfig:correct}
	\includegraphics[width=\textwidth]{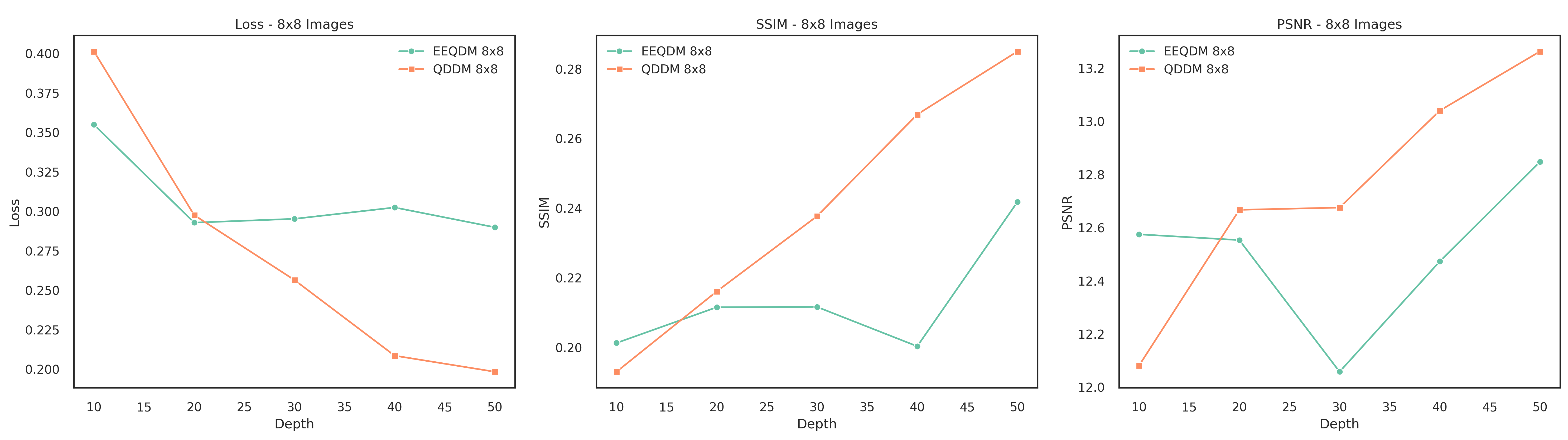} } 

\subfloat[Performance comparison for $16\times16$ MNIST images.]{
	\label{subfig:notwhitelight}
	\includegraphics[width=\textwidth]{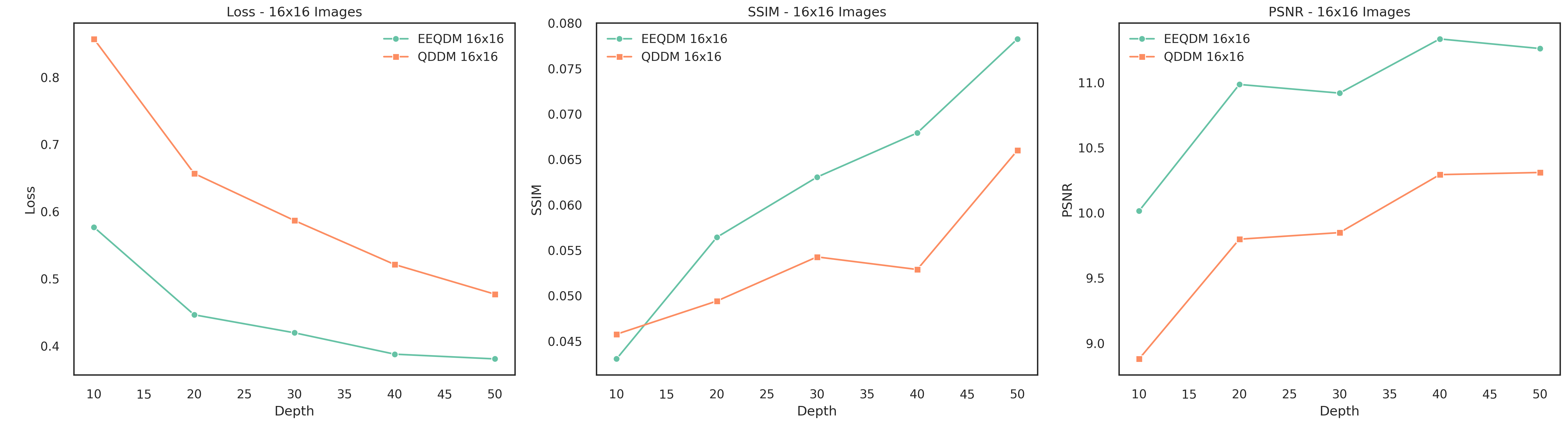} } 
 
\subfloat[Performance comparison for $16\times16$ CIFAR10 images.]{
	\label{subfig:correct}
	\includegraphics[width=\textwidth]{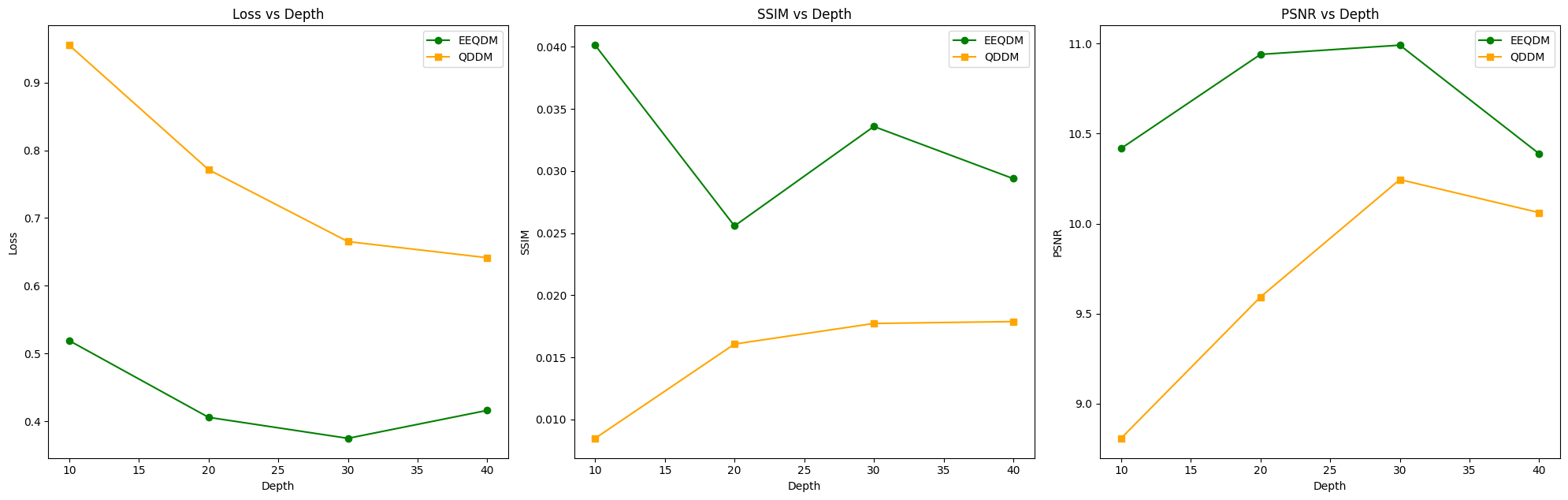} } 

\caption{Performance comparison of EEQDM and QDDM for $8\times8$ and $16\times16$ images across depths from 10 to 50. Results indicate EEQDM's overall superior performance, particularly at greater depths and complex data.}
\label{fig:perf_matrix}

\end{figure}

\begin{itemize}
    \item For 8×8 MNIST images, the performance difference between EEQDM and QDDM is minimal, with QDDM slightly outperforming EEQDM at higher depths in terms of loss values. Both models have comparable execution times, with EEQDM being marginally faster. This indicates that for simpler, smaller-scale tasks, both models perform adequately without significant distinctions.
    \item When processing larger and more complex images, the advantages of EEQDM become evident. For 16×16 MNIST images, EEQDM consistently outperforms QDDM across all metrics, including lower final loss, higher SSIM, and improved PSNR values. This superior performance is achieved with notably faster execution times, especially as model depth increases, suggesting that EEQDM’s entanglement feature not only enhances image reconstruction quality but also improves computational efficiency for higher-resolution grayscale images.
    \item The most significant contrast is observed with 16×16 CIFAR-10 color images. EEQDM demonstrates clear superiority in both performance metrics and computational efficiency. It achieves better image reconstruction quality (evidenced by improved loss, SSIM, and PSNR values) while requiring significantly less execution time compared to QDDM. The computational advantage is particularly pronounced, with EEQDM processing these complex images nearly twice as fast as QDDM at higher depths. This suggests that the pairwise entanglement feature effectively manages the increased computational demands of more complex tasks, offering a better balance between reconstruction quality and processing time.
\end{itemize}

\subsection{Comparison with Classical Denoising Diffusion Model}

\subsubsection{Model Specifications}:
The classical Denoising Diffusion Probabilistic Model (DDPM) is a generative model that incrementally denoises data to produce images. It employs a U-Net architecture with an encoder-decoder design, featuring convolutional layers with ReLU activations. The model includes three depth variants with 2 or 3 layers and different initial channel counts. Designed for $16\times16$ pixel images, the DDPM variants contain between 892 and 6,781 trainable parameters. This classical model serves as a baseline for comparison with our quantum-inspired EEQDM approach, aligning with recent work in quantum-classical comparisons for image generation \cite{koelle2024quantum}.


\subsubsection{Evaluation of Models}: Table \ref{tab:model_comparison} provides a comparison between EEQDM and the classical models. The results reveal intriguing performance trade-offs:

\begin{itemize}
    \item EEQDM demonstrates superior loss reduction as the number of parameters increases, ultimately achieving lower loss with fewer parameters than its classical counterparts. However, this improved performance comes at a significant cost in terms of computation time. EEQDM’s execution time increases exponentially with the parameter count, while classical models maintain relatively constant and much lower execution times regardless of parameter increases.
    \item Despite the classical models’ speed advantage, they show limited improvement in loss reduction even as their complexity grows. This creates a clear trade-off between model accuracy and computational efficiency. EEQDM offers potentially higher accuracy but requires substantially more processing time, particularly for larger parameter sets.
\end{itemize}


\begin{table}[h]
\centering
\begin{tabular}{|l|l|r|r|r|}
\hline
\textbf{Model Type} & \textbf{Depth/Config} & \textbf{Num Params} & \textbf{Final Loss} & \textbf{Execution Time (s)} \\
\hline
EEQDM & 10 & 150 & 0.3674 & 994.36 \\
EEQDM & 20 & 300 & 0.2739 & 1934.08 \\
EEQDM & 30 & 450 & 0.2697 & 2925.98 \\
EEQDM & 40 & 600 & 0.2560 & 3991.76 \\
EEQDM & 50 & 750 & 0.2328 & 5096.88 \\
\hline
Classical & depth2 & 892 & 1.0154 & 175.99 \\
Classical & depth3\_channels2 & 1735 & 1.0058 & 230.34 \\
Classical & depth3\_channels4 & 6781 & 1.0033 & 252.70 \\
\hline
\end{tabular}
\caption{Comparison of EEQDM and Classical Models.}
\label{tab:model_comparison}
\end{table}

\subsection{Comparison with Previous Models on CIFAFR10}

To evaluate the performance of our proposed EEQDM model, we compared it with several state-of-the-art models from previous studies, including U-Net, QU-Net, and Q-Dense. EEQDM exhibits superior performance across all measured metrics while utilizing fewer parameters.

\begin{itemize}
    \item With 750 parameters compared to QDDM’s 1350, EEQDM achieves a 34\% reduction in final loss (0.2993 vs. 0.4536), a 157\% improvement in Structural Similarity Index (SSIM) (0.0433 vs. 0.0169), and a 0.48 dB increase in Peak Signal-to-Noise Ratio (PSNR) (10.65 dB vs. 10.17 dB). Moreover, EEQDM’s execution time is reduced by 35.9\% (5932 seconds vs. 9248 seconds), indicating substantial computational efficiency gains.
    \item The classical model, despite utilizing 6781 parameters, achieves an average loss of 0.9763, whereas EEQDM, with only 750 parameters, attains a final loss of 0.2993—a 69.3\% reduction. The disparity in image quality metrics is even more pronounced: EEQDM’s PSNR of 10.65 dB significantly outperforms the classical model’s -46.77 dB.
\end{itemize}



\subsection{FID Score Comparison of EEQDM and QDDM on MNIST Dataset} The mean FID score for the EEQDM model is 382.36 with a standard deviation of 74.66, indicating moderate variability around the mean. In comparison, the mean FID score for the QDDM model is 420.46 with a lower standard deviation of 44.10, suggesting that the scores are more tightly clustered around the average (see Table \ref{table:fid_comparison} for class-wise comparison).



\begin{table}[h]
    \centering
    \caption{FID Score Comparison for Each Digit.}
    \label{table:fid_comparison}
    \begin{tabular}{|c|c|c|}
        \hline
        Digit & FID\_EEQDM & FID\_QDDM \\
        \hline
        0 & 330.93 & 459.59 \\
        1 & 373.53    & 402.79 \\
        2 & 401.67 & 466.97 \\
        3 & 321.32 & 388.38 \\
        4 & 550.58 & 406.61 \\
        5 & 443.47 & 476.37 \\
        6 & 319.73 & 381.72    \\
        7 & 327.35 & 433.68 \\
        8 & 331.05 & 449.49    \\
        9 & 423.96 & 339.02    \\
        \hline
    \end{tabular}
\end{table}

\section{Conclusion}
This paper introduces the Entanglement-Enhanced Quantum Diffusion Model (EEQDM). It offers clear advantages in processing complex, high-resolution data. EEQDM delivers superior FID, SSIM and PSNR metrics, reduced parameter counts, and faster execution times compared to existing quantum models. Its innovative use of pairwise entanglement is highly effective for intricate data structures, making it more resource-efficient in the quantum domain. An important direction for future work is to develop encoding methods that preserve spatial correlations, enabling correlated data points to be placed within the same entanglement pairs. This approach could enhance model efficiency by leveraging inherent data structures, optimizing quantum resources, and maintaining coherence in entangled states. Moreover, further optimization of quantum circuits is needed to match the speed of classical models. As quantum hardware advances, EEQDM's potential to revolutionize quantum machine learning and complex data processing will grow. Future research should explore EEQDM's applications in other quantum machine learning tasks.

\end{document}